\documentclass[twocolumn,showpacs,aps,pre]{revtex4-1}
\usepackage{amssymb,amsmath}
\usepackage[dvips]{graphicx}

\begin{document}

\title{Spatio-temporal generalization of the Harris criterion and its application to
diffusive disorder}

\author{Thomas Vojta}
\affiliation{Department of Physics, Missouri University of Science and Technology,
Rolla, MO 65409, USA}

\author{Ronald Dickman}
\affiliation{Departamento de  F\'{i}sica, ICEx, Universidade Federal de Minas Gerais,
30123-970, Belo Horizonte - Minas Gerais, Brazil}

\begin{abstract}
We investigate how a clean continuous phase transition is affected by spatio-temporal disorder,
i.e., by an external perturbation that fluctuates in both space and time. We derive a
generalization of the Harris criterion for the stability of the clean critical behavior in
terms of the space-time correlation function of the external perturbation. As an application, we consider
diffusive disorder, i.e., an external perturbation governed by diffusive dynamics, and its effects on
a variety of equilibrium and nonequilibrium critical points. We also discuss the relation
between diffusive disorder and diffusive dynamical degrees of freedom in the example of model C of the
Hohenberg-Halperin classification and comment on Griffiths singularities.
\end{abstract}

\date{\today}

\maketitle

%%%%%%%%%%%%%%%%%%%%%%%%%%%%%%%%%%%%%%%%%%%%%%%%%%%%%%%%%%%%%%%%%%%%%%%%%%%%%%%%%
% Main text starts here
%%%%%%%%%%%%%%%%%%%%%%%%%%%%%%%%%%%%%%%%%%%%%%%%%%%%%%%%%%%%%%%%%%%%%%%%%%%%%%%%
\section{Introduction}
\label{sec:Intro}
%%%%%%%%%%%%%%%%%%%%%%%%%%%%%%%%%%%%%%%%%%%%%%%%%%%%%%%%%%%%%%%%%%%%%%%%%%%%%%%%%

If a many-particle system undergoes a continuous phase transition, one can ask whether
or not weak external perturbations destabilize the critical behavior, i.e., whether or
not they change the universality class of the phase transition. Such perturbations could,
for example, stem from impurities, defects and other types of spatial disorder.
They could also stem from external temporal fluctuations or from the coupling to
more complicated external degrees of freedom.

Harris \cite{Harris74} investigated how a given clean critical point is affected by
time-independent uncorrelated
spatial disorder that locally favors one phase over the other but does not break
any of the order parameter symmetries (so-called random-mass or random-$T_c$ disorder).
Using a beautiful heuristic argument, he derived a criterion for the stability of the
clean critical behavior: If the spatial correlation length exponent $\nu$ of
a $d$-dimensional clean system fulfills the inequality $d\nu>2$,
weak disorder is irrelevant and does not change the critical behavior.
If $d\nu<2$, disorder is relevant, and the character of the transition must change.
Note that the Harris criterion does \emph{not} determine the ultimate fate of the transition
in the case $d\nu<2$; it could be as simple as a new set of critical
exponents \footnote{The correlation length exponent of the disordered system
must then fulfill the inequality $d\nu>2$ \cite{CCFS86}.}, or conventional
power-law scaling could be replaced by activated scaling, or the transition could be
completely destroyed by smearing (for reviews see, e.g., \cite{Vojta06,Vojta10}).
Note, however, that the Harris criterion does determine the character of
quantum Griffiths singularities (if any) at a disordered quantum phase transition
\cite{VojtaHoyos14}.

Over the years, the Harris criterion has been generalized in several directions.
Weinrib and Halperin \cite{WeinribHalperin83} investigated long-range correlated spatial disorder
characterized by a correlation function that decays as $|\mathbf{x}|^{-a}$ with
distance $|\mathbf{x}|$. If $a>d$, the stability of the clean critical behavior is controlled
by the usual Harris criterion $d\nu>2$; but for $a<d$, the inequality gets replaced
by $a\nu > 2$. Luck \cite{Luck93a} formulated the criterion in terms of the wandering
exponent $\omega$ that characterizes the fluctuations of an arbitrary spatial modulation.

All perturbations mentioned so far involve fluctuations in space, but spatially uniform temporal fluctuations
can be studied as well. Kinzel \cite{Kinzel85} showed that uncorrelated temporal disorder
destabilizes a nonequilibrium phase transition if the correlation time exponent
$\nu_\parallel=z\nu$ violates the inequality $z\nu > 2$ (here, $z$ is the dynamical
critical exponent); the same criterion was put forward by Alonso and Mu\~noz \cite{AlonsoMunoz01}
for Ising models. What about perturbations that
fluctuate in both space and time and are characterized by nontrivial space-time correlations?

To answer this question, we derive in this paper a generalization of the Harris criterion to arbitrary spatio-temporal
disorder of random-mass type. We then apply this criterion to the important case of diffusive disorder.
Our paper is organized as follows. In Sec.\ \ref{sec:Criterion}, we derive the general stability
criterion and show that all the criteria mentioned above can be viewed as special cases of this
criterion. In Sec.\ \ref{sec:Diffusion}, we focus on diffusive disorder and work out the scaling
of the disorder fluctuations in this case. Section \ref{sec:Applications} is devoted to the
application of our criterion to several equilibrium and nonequilibrium phase transitions.
We conclude in Sec.\ \ref{sec:Conclusions}.

%%%%%%%%%%%%%%%%%%%%%%%%%%%%%%%%%%%%%%%%%%%%%%%%%%%%%%%%%%%%%%%%%%%%%%%%%%%%%%%%
\section{General stability criterion}
\label{sec:Criterion}
%%%%%%%%%%%%%%%%%%%%%%%%%%%%%%%%%%%%%%%%%%%%%%%%%%%%%%%%%%%%%%%%%%%%%%%%%%%%%%%%%
\subsection{Basic formalism}
%%%%%%%%%%%%%%%%%%%%%%%%%%%%%%%%%%%%%%%%%%%%%%%%%%%%%%%%%%%%%%%%%%%%%%%%%%%%%%%%%

In this section we derive a criterion for the stability of a clean critical point
against general spatio-temporal random-mass disorder. Let us start from a clean,
(translationally invariant in space and time) equilibrium or nonequilibrium system
that undergoes a continuous phase transition characterized by a set of critical
exponents. We introduce spatio-temporal disorder by making the local distance
from criticality $r$ a random function of position $\mathbf{x}$ and time $t$,
\begin{equation}
r \to r_0 + w \, n(\mathbf{x},t)
\label{eq:r}
\end{equation}
where $w$ is the disorder amplitude, and the (random) field $n(x,t)$ describes its
space and time dependencies. In a
lattice model, this type of disorder could be achieved, e.g., by having bond
strengths that vary with $\mathbf{x}$ and $t$.
We emphasize that $n(\mathbf{x},t)$ is an \emph{external} perturbation rather
than a system degree of freedom. This means, there is no feedback from the system
on $n(x,t)$. We will come back to this question in Sec.\
\ref{sec:Applications}.
Without loss of generality, we can assume that $n(\mathbf{x},t)$ has zero
average (a nonzero average can be absorbed into $r_0$),
\begin{equation}
[n(\mathbf{x},t)]_\textrm{dis} = 0~.
\label{eq:nav}
\end{equation}
It is characterized by a correlation function which we assume to be translationally
invariant,
\begin{equation}
[n(\mathbf{x},t)\,n(\mathbf{x'},t')]_\textrm{dis} = G_{nn}(\mathbf{x}-\mathbf{x'},t-t')~.
\label{eq:Gnn}
\end{equation}
Here, $[\ldots]_\textrm{dis}$ denotes the average over the disorder distribution.

The basic idea underlying the stability criterion is to compare the fluctuations
of the local distance from criticality with the global distance from criticality.
Close to a critical point, the system effectively averages over lengths of the order
of the correlation length $\xi$ and times of the order of the correlation
time $\xi_t$. We therefore need to average the local distance
from criticality (\ref{eq:r}) over a $(d+1)$-dimensional correlation volume $V_\xi$ of size
$\xi^d \times \xi_t$, giving $\bar r(\xi,\xi_t) = r_0 + w \bar n(\xi,\xi_t)$ with
\begin{equation}
\bar n(\xi,\xi_t)  = \frac {1}{\xi^d \xi_t} \int_{V_\xi} d^d x dt\, n(\mathbf{x},t)~.
\label{eq:nbar}
\end{equation}
The disorder average of $\bar n$ obviously vanishes, and its variance is given by
\begin{eqnarray}
\label{eq:sigmanbar}
\sigma^2_{\bar n}(\xi,\xi_t) &=& [\bar n^2(\xi,\xi_t)]_\textrm{dis} \\
                             &=& \frac {1}{\xi^{2d}\xi_t^2}\int d^dxd^dx'dtdt' \, [n(\mathbf{x},t)n(\mathbf{x'},t')]_\textrm{dis}~\nonumber \\
                             &=& \frac {1}{\xi^{2d}\xi_t^2}\int d^dxd^dx'dtdt' \,  G_{nn}(\mathbf{x}-\mathbf{x'},t-t')~.~ \nonumber
\end{eqnarray}
Using the translational invariance of the correlation function, we can carry out one set of space-time integrations
and approximate $\sigma^2_{\bar n}(\xi,\xi_t)$ by
\begin{equation}
\sigma^2_{\bar n}(\xi,\xi_t) \approx \frac {1}{\xi^{d}\xi_t}\int_{-\xi/2}^{\xi/2} d^dx \int_{-\xi_t/2}^{\xi_t/2}dt \,  G_{nn}(\mathbf{x},t)~.~
\label{eq:sigmanbar_approx}
\end{equation}
This approximation correctly captures the leading $\xi$ and $\xi_t$ dependencies of the variance. The boundary
conditions of the correlation volume are not treated correctly, but this is unimportant for our purposes.
The quantity $\sigma_{\bar r}(\xi,\xi_t) = w \sigma_{\bar n}(\xi,\xi_t)$ characterizes the fluctuations of
the local distance from criticality between different correlation volumes.

To assess the stability of the clean critical behavior, we now compare $\sigma_{\bar r}(\xi,\xi_t)$ with
the global distance from criticality $r_0 \sim \xi^{-1/\nu}$. If $\sigma_{\bar r} / r_0 \to 0$ as the critical
point is approached assuming the clean critical behavior, the disorder becomes less and less
important and the system is asymptotically clean. In this case the clean critical behavior is (perturbatively)
stable against the disorder.
In contrast, if assuming the clean critical exponents implies that $\sigma_{\bar r} / r_0 \to \infty$ for $r_0\to 0$,
a homogenous transition with the clean behavior is impossible (as different correlation volumes would end
up on  different sides of the critical point). The clean critical point is therefore unstable.

Consequently, the general criterion reads: The clean critical behavior is (perturbatively) stable against weak disorder,
if
\begin{equation}
\xi^{2/\nu -d}\xi_t^{-1} \int_{-\xi/2}^{\xi/2} d^dx \int_{-\xi_t/2}^{\xi_t/2}dt \,  G_{nn}(\mathbf{x},t) \to 0
\label{eq:criterion}
\end{equation}
as the critical point is approached, i.e, for $\xi, \xi_t \to \infty$ with the appropriate scaling relation
between $\xi$ and $\xi_t$. (For conventional power-law dynamical scaling this means $\xi_t \sim \xi^z$.)

%%%%%%%%%%%%%%%%%%%%%%%%%%%%%%%%%%%%%%%%%%%%%%%%%%%%%%%%%%%%%%%%%%%%%%%%%%%%%%%%%
\subsection{Simple examples}
%%%%%%%%%%%%%%%%%%%%%%%%%%%%%%%%%%%%%%%%%%%%%%%%%%%%%%%%%%%%%%%%%%%%%%%%%%%%%%%%%

In this subsection, we work out the stability criterion for several simple examples of
disorder correlation functions. In this way, we can rederive the criteria discussed in
Sec.\ \ref{sec:Intro} as special cases of our theory.

\paragraph{Uncorrelated spatial disorder.} If the disorder is uncorrelated in space and
time-independent (i.e., perfectly correlated in time), the disorder correlation function
reads $G_{nn}(\mathbf{x},t) \sim \delta(\mathbf{x})$. Carrying out the integral
(\ref{eq:sigmanbar_approx}) gives $\sigma^2_{\bar n}(\xi,\xi_t) \sim \xi^{-d}$ in agreement
with the central limit theorem. The clean critical behavior is stable if
$\sigma_{\bar r} / r_0  \sim \xi^{-d/2+1/\nu} \to 0$ for $\xi\to \infty$.
This implies the exponent inequality
\begin{equation}
d\nu > 2~.
\label{eq:Harris}
\end{equation}
We thus recover the original Harris criterion \cite{Harris74}.

\paragraph{Long-range correlated spatial disorder.} The disorder correlation
function is time-independent and behaves as
$G_{nn}(\mathbf{x},t) \sim |\mathbf{x}|^{-a}$ for large $|\mathbf{x}|$.
If we carry out the integral (\ref{eq:sigmanbar_approx}), we find
$\sigma^2_{\bar n}(\xi,\xi_t) \sim \xi^{-d}$ for $a>d$ but
$\sigma^2_{\bar n}(\xi,\xi_t) \sim \xi^{-a}$ for $a<d$. For $a<d$,
the clean critical point is therefore stable if
\begin{equation}
a\nu > 2
\label{eq:WeinribHalperin}
\end{equation}
in agreement with Weinrib and Halperin \cite{WeinribHalperin83},
while the normal Harris criterion governs the case $a>d$.

\paragraph{Uncorrelated temporal disorder.} For uncorrelated purely temporal (i.e.,
space-independent)
disorder, the correlation function is given by  $G_{nn}(\mathbf{x},t) \sim \delta(t)$.
The integral (\ref{eq:sigmanbar_approx}) results in
$\sigma^2_{\bar n}(\xi,\xi_t) \sim \xi_t^{-1} \sim \xi^{-z}$. The condition
$\sigma_{\bar r} / r_0 \to 0$ as $\xi \to \infty$ then implies, that
the clean critical behavior is stable if
\begin{equation}
z\nu > 2
\label{eq:Kinzel}
\end{equation}
as stated by Kinzel \cite{Kinzel85} as well as Alonso and Mu\~noz \cite{AlonsoMunoz01}.

\paragraph{Long-range correlated temporal disorder.} If the disorder is
purely temporal and characterized by power-law correlations
$G_{nn}(\mathbf{x},t) \sim |t|^{-a}$ for large $|t|$, we find
$\sigma^2_{\bar n}(\xi,\xi_t) \sim \xi_t^{-1}$ for $a>1$ but
$\sigma^2_{\bar n}(\xi,\xi_t) \sim \xi_t^{-a}$ for $a<1$. The stability
criterion thus reads
\begin{equation}
a z\nu > 2
\label{eq:Kinzel_lr}
\end{equation}
for $a<1$ while the case $a>1$ is governed by eq.\ (\ref{eq:Kinzel}).

\paragraph{Uncorrelated spatio-temporal disorder.} If the disorder
is uncorrelated in both space and time,  $G_{nn}(\mathbf{x},t) \sim \delta(\mathbf{x})\delta(t)$,
the variance (\ref{eq:sigmanbar_approx}) of the local distance from
criticality reads $\sigma^2_{\bar n}(\xi,\xi_t) \sim \xi^{-d}\xi_t^{-1} \sim \xi^{-(d+z)}$.
As stated in Ref.\ \cite{AlonsoMunoz01} (for Ising models), the clean critical behavior is therefore
stable if
\begin{equation}
(d+z)\nu > 2~.
\label{eq:space+time}
\end{equation}

%%%%%%%%%%%%%%%%%%%%%%%%%%%%%%%%%%%%%%%%%%%%%%%%%%%%%%%%%%%%%%%%%%%%%%%%%%%%%%%%
\section{Diffusive disorder}
\label{sec:Diffusion}
%%%%%%%%%%%%%%%%%%%%%%%%%%%%%%%%%%%%%%%%%%%%%%%%%%%%%%%%%%%%%%%%%%%%%%%%%%%%%%%%%

We now turn to our main topic, the effects  of diffusive disorder on a clean critical point.
In this case, the dynamics of the disorder field $n(\mathbf{x},t)$ can be described
by the Langevin equation
\begin{equation}
\frac {\partial}{\partial t} n(\mathbf{x},t) = D \nabla^2 n(\mathbf{x},t) + \zeta(\mathbf{x},t)
\label{eq:Langevin}
\end{equation}
where $D$ is the diffusion constant and $\zeta(\mathbf{x},t)$ is a conserving noise.
In order to derive our stability criterion, we need the correlation function $G_{nn}(\mathbf{x},t)$
of the diffusive field. It can be determined using standard techniques, as will be sketched
in Appendix \ref{sec:Appendix_A}. We find
\begin{equation}
G_{nn}(\mathbf{x},t) = \frac {A}{(4\pi D|t|)^{d/2}} \exp\left[- \frac{\mathbf{x}^2}{4 D |t|} \right]
\label{eq:Gnn_diffusion}
\end{equation}
where $A$ is some constant. If the field $n(\mathbf{x},t)$ is in equilibrium, $A$ can be
expressed in terms of the compressibility and the temperature, $A=k_B T (\partial n/\partial \mu$).

We proceed by considering the average $\bar n(\xi,\xi_t)$ of the diffusive field over a correlation
volume. Its variance can be estimated using eq.\ (\ref{eq:sigmanbar_approx}). We first carry out the
$x$-integration, distinguishing two regimes. For early times, $4Dt < (\xi/2)^2$, the $x$-integration can be extended
to infinity, giving $\int d^dx \, G_{nn}(\mathbf{x},t) = A$. For late times, $4Dt > (\xi/2)^2$,
the exponential in $G_{nn}$ is approximately equal to unity. The $x$-integration
in eq.\ (\ref{eq:sigmanbar_approx}) thus yields $\int d^dx \, G_{nn}(\mathbf{x},t) =
A\xi^d/(4 \pi D |t|)^{d/2}$.

To perform the remaining time integration in  eq.\ (\ref{eq:sigmanbar_approx}),
we need to distinguish the cases $2 D\xi_t < (\xi/2)^2$ and $2D\xi_t > (\xi/2)^2$.
In the former case, we can use the above early-time result for all $t$. For  $2D\xi_t < (\xi/2)^2$, we therefore
obtain
\begin{equation}
\sigma^2_{\bar n}(\xi,\xi_t) \approx \frac {1}{\xi^{d}\xi_t} \int_{-\xi_t/2}^{\xi_t/2}dt \, A = \frac {A}{\xi^d}~,~
\label{eq:sigma_diffusive_early}
\end{equation}
independent of $\xi_t$, i.e., the same behavior as for uncorrelated purely spatial disorder.

The case  $2D\xi_t > (\xi/2)^2$ is more complicated because the time integration range covers
both the early-time and the late-time regimes of the $x$-integration above. We therefore
split the time integration range into two intervals, $0< 4Dt<(\xi/2)^2$ and $(\xi/2)^2 < 4Dt <2D \xi_t$.
The evaluation of the resulting integrals is straight forward (see Appendix \ref{sec:Appendix_B}) and yields the following
leading behavior for $2D\xi_t > (\xi/2)^2$:
\begin{equation}
\sigma^2_{\bar n}(\xi,\xi_t) \sim \left\{
\begin{array}{ll}
A D^{-1} \xi_t^{-1}\xi^{2-d}                  & (d>2)  \\
A D^{-1} \xi_t^{-1} \ln (8D\xi_t/\xi^2) \quad & (d=2)  \\
A D^{-d/2} \xi_t^{-d/2}                         & (d<2)
\end{array}
\right. ~.
\label{eq:sigma_diffusive_late}
\end{equation}
Interestingly, the variance is independent of $\xi$ for $d<2$.
Comparing eqs.\ (\ref{eq:sigma_diffusive_early}) and (\ref{eq:sigma_diffusive_late}), we see that
the fluctuations $\sigma^2_{\bar n}(\xi,\xi_t)$  in the short-time case
(\ref{eq:sigma_diffusive_early}) are larger than those in the long-time case (\ref{eq:sigma_diffusive_late}).
This is caused by the extra averaging in time direction that happens in the long-time case.

\begin{figure}
\includegraphics[width=8.3cm]{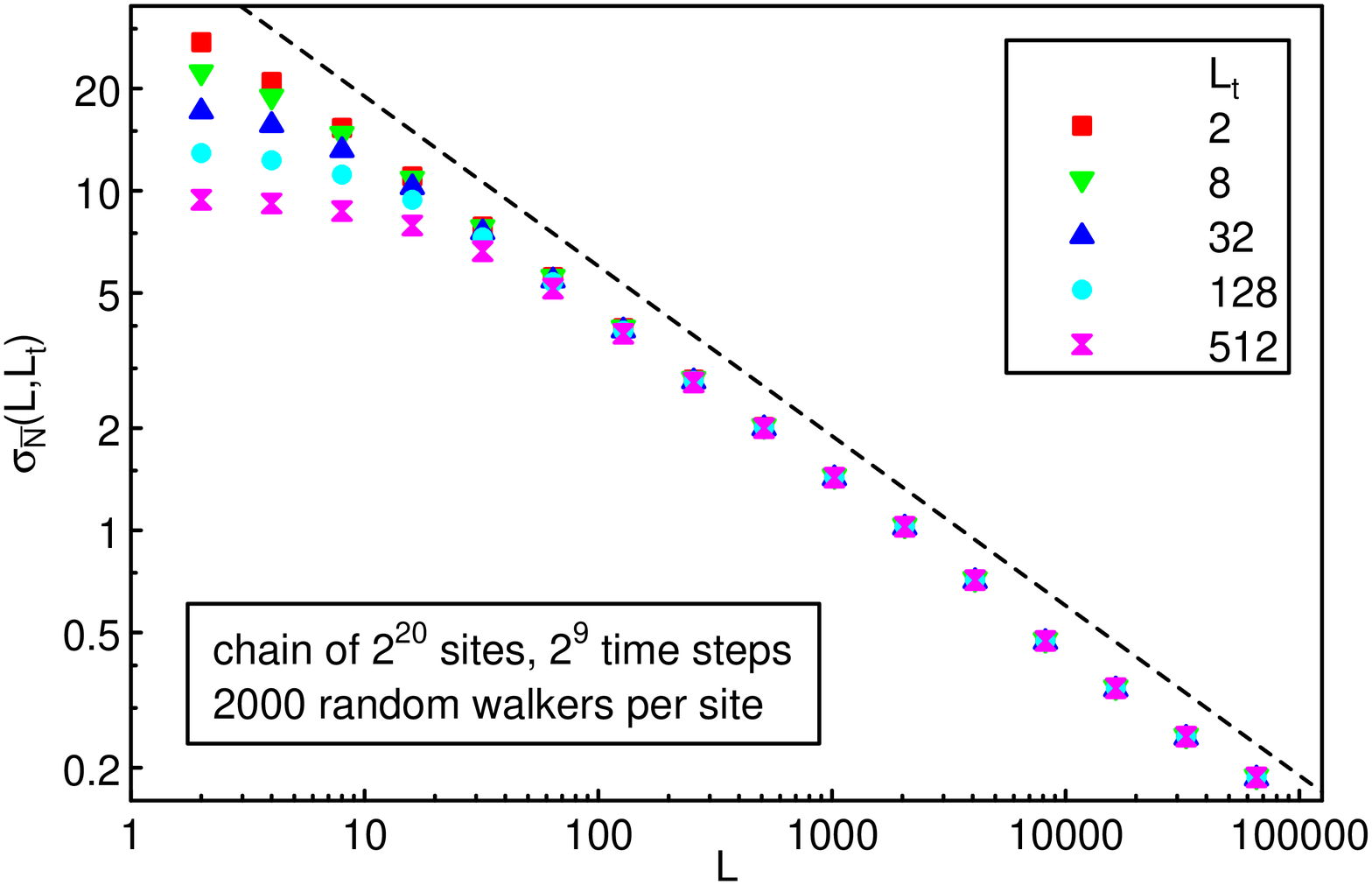}
\includegraphics[width=8.3cm]{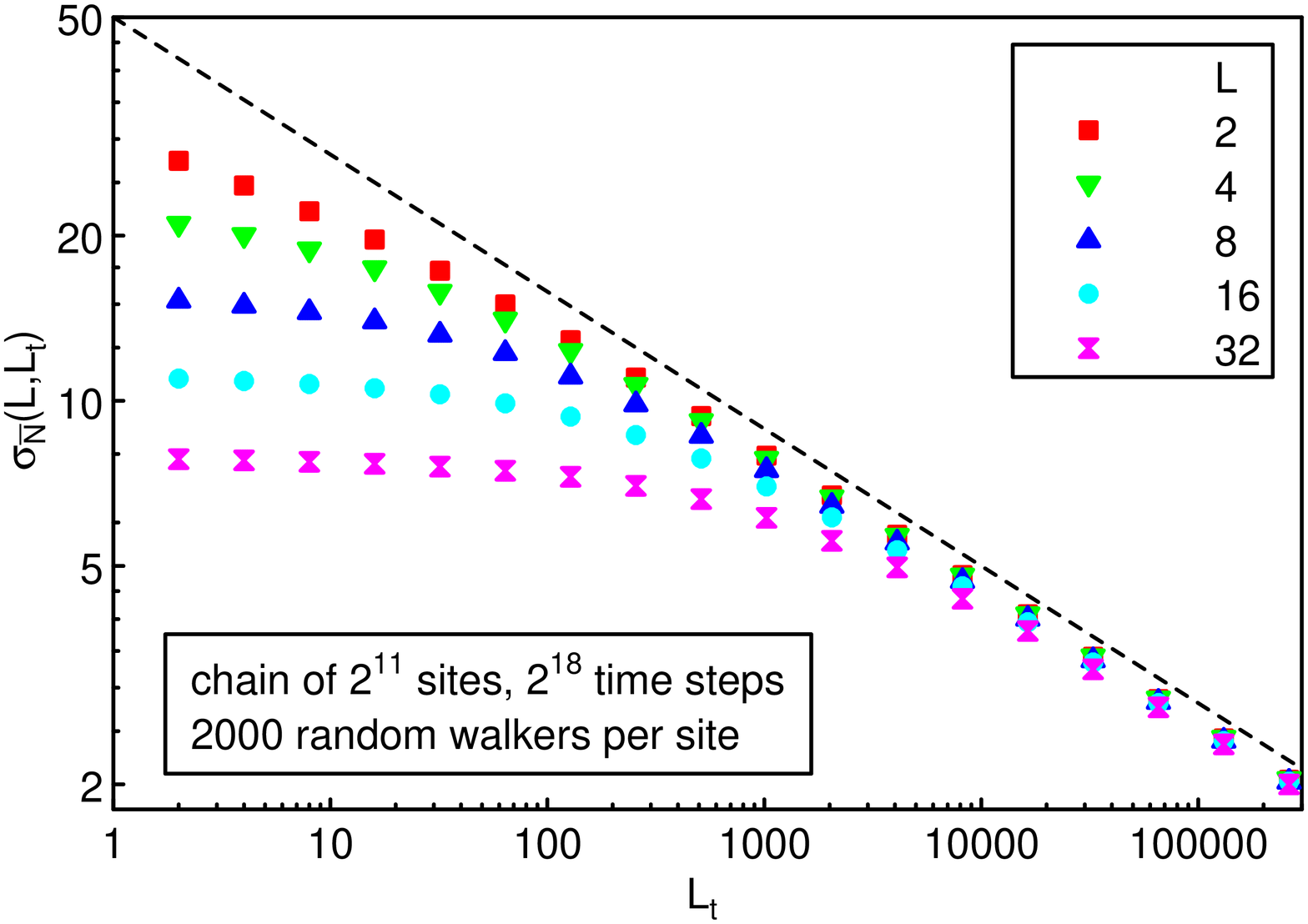}
\caption{(color online) Standard deviation $\sigma_{\bar N}$ of the number of walkers $N(i,t)$ averaged
over a space-time volume of length $L$ and time-length $L_t$. The dashed lines represent
power laws with exponent $-1/2$ (upper panel) and $-1/4$ (lower panel) and arbitrary prefactors.}
\label{fig:sig}
\end{figure}

To test the predictions (\ref{eq:sigma_diffusive_early}) and (\ref{eq:sigma_diffusive_late}), we performed
computer simulations of random walkers in one dimension. Initially, a large number of walkers are placed at random
on the sites of a one dimensional chain. Each walker then performs an unbiased random walk, i.e., in each time step,
it hops left or right with equal probability. The number of walkers $N(i,t)$ occupying site $i$ at time $t$
is a realization of our diffusive field. Figure \ref{fig:sig} shows the fluctuations of $\bar N(L,L_t)$ which is the average of  $N(i,t)$
over a space-time volume of length $L$ and time-length $L_t$.
The data in the upper panel agree with eq.\ (\ref{eq:sigma_diffusive_early}) while the
lower panel confirms eq.\ (\ref{eq:sigma_diffusive_late}) in the one-dimensional case.

To assess the stability of a given clean critical point, we now evaluate the ratio
$\sigma_{\bar r} / r_0 = w\sigma_{\bar n} / r_0$. This analysis depends on the value
of the dynamical critical exponent $z$. If $z<2$, the correlation time behaves as
$\xi_t \sim \xi^z < \xi^2$ for $\xi\to \infty$. Asymptotically, the diffusive disorder in a correlation volume
is thus in the static limit in which the fluctuations are given by eq.\ (\ref{eq:sigma_diffusive_early}).
Consequently, $\sigma_{\bar r} / r_0  \sim \xi^{-d/2+1/\nu}$ for $\xi\to \infty$,
implying that the stability against diffusive disorder
for $z<2$ is controlled by the normal Harris criterion
\begin{equation}
d\nu > 2~.
\label{eq:Harris_2}
\end{equation}

In contrast, for $z>2$, we have $\xi_t \sim \xi^z > \xi^2$ for $\xi\to \infty$
which means that the diffusive disorder in a correlation volume is in the fluctuating
limit in which the variance is given by eq.\ (\ref{eq:sigma_diffusive_late}).
Evaluating the ratio  $\sigma_{\bar r} / r_0$ as before, we find that the clean critical
point with $z>2$ is stable against diffusive disorder if
\begin{eqnarray}
(d+z-2)\nu &>& 2    \qquad\qquad (d>2),\label{eq:criterion_d>2}\\
z\nu &>& 2          \qquad\qquad (d=2),\label{eq:criterion_d=2}\\
dz\nu &>& 4         \qquad\qquad (d<2).\label{eq:criterion_d<2}
\end{eqnarray}
These inequalities are less stringent that the normal Harris criterion, in
agreement with the suppression of the fluctuations discussed after eq.\
(\ref{eq:sigma_diffusive_late}).

The results of this section can be summarized as follows: For critical points with
$z<2$, diffusive disorder is as relevant as uncorrelated spatial (static)
disorder. (Note that this does \emph{not} mean that diffusive disorder and
 spatial disorder necessarily lead to the same ultimate fate of the transition.)
For critical points with $z>2$, by contrast, diffusive disorder is
less relevant than uncorrelated spatial disorder.

%%%%%%%%%%%%%%%%%%%%%%%%%%%%%%%%%%%%%%%%%%%%%%%%%%%%%%%%%%%%%%%%%%%%%%%%%%%%%%%%
\section{Applications}
\label{sec:Applications}
%%%%%%%%%%%%%%%%%%%%%%%%%%%%%%%%%%%%%%%%%%%%%%%%%%%%%%%%%%%%%%%%%%%%%%%%%%%%%%%%%

We now apply the diffusive disorder stability criterion to a number of equilibrium and nonequilibrium
phase transitions.

\subsection{Absorbing state transitions}

The estimates for the (clean) critical exponents of the transitions discussed in this subsection
are taken from Refs.\ \cite{Hinrichsen00,Odor04} and references therein.

\setcounter{paragraph}{0}
\paragraph{Directed percolation.} In the directed percolation universality class, the dynamical critical exponent
approximately takes the values $z=\nu_\parallel/\nu \approx 1.58$, 1.76, and 1.90 in one, two, and
three space dimensions, respectively. As $z<2$ in all these dimensions, the diffusive disorder is
asymptotically in the static limit, and its relevance is governed by eq.\
(\ref{eq:Harris_2}), i.e., by the normal Harris criterion. The spatial correlation
length exponent takes the values $\nu \approx 1.097$, 0.73, and 0.58. Harris' inequality
$d\nu > 2$ is thus violated in all dimensions implying that the directed percolation
universality class is unstable against diffusive disorder.

This agrees with explicit results for specific models. Dickman \cite{Dickman09} studied
a version of the one-dimensional contact process in which diffusing impurities lead to
space and time dependent infection rates. He found a continuous phase transition whose
critical behavior differs from the directed percolation universality class.
Note that the physics of diffusing impurities differs from that of diffusing active and inactive sites.
As the densities of active and inactive sites are not conserved, the latter case does not lead to
non-trivial space-time correlations, and the critical behavior remains in the directed percolation
universality class \cite{JensenDickman93}.
% ???Other examples, Malaria model???

\paragraph{Parity-conserving class in 1D.} Transitions in the one-dimensional parity-conserving
universality class have a dynamical exponent $z \approx 1.76 <2$.
Thus, the stability of the clean critical point is governed by  eq.\ (\ref{eq:Harris_2}) because the diffusive disorder
is asymptotically in the static limit. As $\nu\approx 1.83$, Harris' inequality
$d\nu > 2$ is violated, and the clean critical behavior is unstable against diffusive disorder.

\paragraph{Voter model class in 2D.} The voter model (or DP2) class in two space dimensions
has a dynamical exponent $z=2$, putting the diffusive disorder
right at the boundary between the static and fluctuating limits where eqs.\   (\ref{eq:Harris_2})
and (\ref{eq:criterion_d=2}) coincide. As $\nu=1/2$, Harris' inequality
$d\nu > 2$ is violated, and diffusive disorder is relevant.

\paragraph{Tricritical directed percolation.} In two space dimensions, the tricritical directed
percolation universality class features a dynamical critical exponent of $z\approx 2.11 > 2$.
In contrast to the examples above, the relevance of diffusive disorder is therefore not
governed by Harris' inequality but by the new criterion (\ref{eq:criterion_d=2}),
$z\nu > 2$. As $\nu\approx 0.547$, this criterion is violated. This means that the tricritical directed
percolation universality class in two space dimensions is unstable against diffusive disorder.
The same holds for three and higher dimensions in which the clean critical behavior is of mean-field
type with $z=2$ and $\nu=1/2$.

\subsection{Kinetic Ising and Heisenberg magnets}
\label{subsec:Ising}

As examples of equilibrium phase transitions, we consider kinetic Ising and Heisenberg models
with purely relaxational dynamics (model A of the Hohenberg-Halperin classification \cite{HohenbergHalperin77}).
Microscopically, this dynamics can be realized, e.g., by the Glauber or Metropolis algorithms \cite{MRRT53,Glauber63}.

\setcounter{paragraph}{0}
\paragraph{2D Ising model.} The dynamical critical exponent of the two-dimensional kinetic Ising model takes the
value $z\approx 2.17$ (see, e.g., Ref.\ \cite{NightingaleBlote00}). The relevance of diffusive disorder
is thus controlled by the new criterion (\ref{eq:criterion_d=2}), $z\nu > 2$.
As $\nu=1$ for the two-dimensional Ising model, this criterion is fulfilled. Consequently, the clean critical
behavior is stable against diffusive disorder.

\paragraph{3D Ising model.} The critical point of the three-dimensional Ising model with purely relaxational dynamics features a dynamical
critical exponent of $z\approx 2.04$ \cite{WanslebenLandau91,*Grassberger95,*JMSZ99}.
Its stability against diffusive disorder is therefore governed by the criterion (\ref{eq:criterion_d>2}),
$(d+z-2)\nu > 2$.  The correlation length exponent reads $\nu\approx 0.630$ \cite{DengBlote03,Hasenbusch10}.
Thus,  $(d+z-2)\nu \approx 1.92 <2$ implying that diffusive disorder is a relevant perturbation.

\paragraph{3D Heisenberg model.} In contrast to the kinetic Ising model, the dynamical exponent of the
Heisenberg model with relaxational dynamics is actually below 2. Monte-Carlo estimates give
$z\approx 1.97$ \cite{PeczakLandau93,FernandesdaSilvaDrugowich06}. The relevance of diffusive
disorder is therefore governed by Harris' inequality $d\nu >2$. The correlation length exponent
takes the value $\nu \approx 0.711$ \cite{CHPRV02}. Diffusive disorder is therefore irrelevant.

It is interesting to compare the effects of diffusive disorder (as considered here) with the coupling
of the order parameter to a diffusive field that is \emph{in equilibrium} with the rest of the system.
The latter case corresponds to model C of the Hohenberg-Halperin classification. In general, the physics of the two
cases is different because the diffusive disorder is externally given and not influenced by the
system itself. In model C, by contrast, order parameter and diffusive field mutually influence each other.
However, if we only ask whether or not a given critical behavior is stable against a weak coupling to either
diffusive disorder or a diffusive dynamical field, the two cases are actually equivalent. In renormalization
group language, our generalized stability criterion tests whether the (tree-level) scale dimension at the
clean critical point of the coupling between the order parameter and the diffusive disorder/field is
positive or negative. This tree-level scale dimension is the same for externally given disorder and a
dynamical field. This becomes particulary obvious within the replica formalism (see, e.g., Ref.\ \cite{Cardy_book96})
where the only difference between the two cases is in the replica structure of the perturbing term which does not play
a role at tree level.

These arguments suggest that our generalized criterion, eqs.\ (\ref{eq:Harris_2}) to (\ref{eq:criterion_d<2}),
controls not only the stability against weak diffusive disorder but also the stability against weak coupling to a diffusive
dynamic field. Recently, the effects of a diffusive dynamic field on the model-A phase transition were
studied using a functional renormalization group in $2 \le d \le 4$ \cite{MSPB13}. The authors found several regimes depending
on $d$ and the number of order parameter components. The boundary between the regime in which $z$
remains at its model-A value and the regime where it changes was determined to be given by $\alpha/\nu=z-2$. Using the scaling relation
$2-\alpha = d\nu$ , this is exactly equivalent to the condition $(d+z-2)\nu=2$, in agreement with our
eq.\ (\ref{eq:criterion_d>2}).

%%%%%%%%%%%%%%%%%%%%%%%%%%%%%%%%%%%%%%%%%%%%%%%%%%%%%%%%%%%%%%%%%%%%%%%%%%%%%%%%%
\section{Conclusions}
\label{sec:Conclusions}
%%%%%%%%%%%%%%%%%%%%%%%%%%%%%%%%%%%%%%%%%%%%%%%%%%%%%%%%%%%%%%%%%%%%%%%%%%%%%%%%%

In summary, we have studied the stability of critical points against general spatio-temporal disorder of
random mass type, i.e., disorder that changes the local distance from criticality but does not break
any order-parameter symmetry. By analyzing the relative fluctuations of the distance from criticality
of a (space-time) correlation volume, we have derived a generalization of the Harris criterion in terms of the
space-time correlation function of the disorder. The original Harris criterion \cite{Harris74} for
uncorrelated spatial disorder, Weinrib and Halperin's version \cite{WeinribHalperin83}
for power-law correlated spatial disorder, as well as Kinzel's criterion \cite{Kinzel85} for
uncorrelated temporal disorder emerge as special cases of our theory.

We have focused on the important case of diffusive disorder in which the local distance from criticality
is modulated by a diffusive density $n(\mathbf{x},t)$. In this case, the form of the stability criterion
depends on the value of the (clean) dynamical exponent $z$. If $z<2$, the correlation time $\xi_t$ grows more slowly
than $\xi^2$ as the critical point is approached. Consequently, the diffusive disorder is
asymptotically in the static limit, and its relevance is governed by the normal Harris criterion.
For $z>2$, the disorder is less relevant because there is additional averaging in time direction.
The resulting stability criterion is given in eqs.\ (\ref{eq:criterion_d>2}) to (\ref{eq:criterion_d<2}).
We have used this criterion to predict the effects of diffusive disorder on a number of
equilibrium and nonequilibrium phase transitions and to organize existing results.

Our generalized stability criterion governs the influence of \emph{weak} diffusive disorder. What about
rare strong disorder fluctuations and the rare regions that support them? Specifically, can diffusive
disorder lead to power-law Griffiths singularities analogous to those caused by spatial disorder
in certain nonequilibrium and quantum phase transitions (see, e.g., Ref.\ \cite{Vojta06})? Power-law Griffiths
singularities arise because the time scale associated with the order parameter fluctuations on a rare region grows exponentially with
its linear size $L$. In the case of diffusive disorder, however, a rare disorder fluctuation of size $L$
has a finite lifetime itself: it increases only as $L^2$ with the size of the region. Therefore,
the lifetime of the disorder fluctuations is much too short to support power-law Griffiths
singularities. In agreement with this argument, Griffiths singularities were not observed in the
simulations of the contact process with mobile disorder \cite{Dickman09} while static spatial disorder
does lead to Griffiths singularities
\cite{Noest86,HooyberghsIgloiVanderzande03,*HooyberghsIgloiVanderzande04,VojtaDickison05,*VojtaFarquharMast09,*Vojta12}.

We emphasize that we have considered diffusive \emph{disorder} which is externally given and not influenced
by the system itself. This needs to be distinguished from the case in which a diffusive dynamic degree
of freedom and the order parameter mutually influence each other. An example of the latter situation is
model C of the Hohenberg-Halperin classification where the order parameter and the diffusive field are
in equilibrium with each other. While the physics of diffusive disorder and a diffusive dynamic degree
of freedom are generally different, the renormalization group arguments laid out at the end of Sec.\
\ref{subsec:Ising} suggest that the stability of a critical point against both types of perturbations
is governed by the same criteria. Indeed, our stability criterion (\ref{eq:criterion_d>2}) for diffusive
disorder agrees with the corresponding boundary for the stability of model-A critical behavior against
coupling to a diffusive dynamic field in model C \cite{MSPB13}.

A criterion similar to the one derived here was recently used to show that
particle density fluctuations in a conserved stochastic sandpile destabilize the directed
percolation critical behavior \cite{DickmandaCunha15}.
Note however, that in this system, the coupling between the conserved particle
density and the order parameter is \emph{not} weak. As a result, density fluctuations grow more
slowly than those of a diffusive field, leading to ``hyperuniformity'' \cite{HexnerLevine15} in sandpiles.

It is also interesting to consider the effects of spatio-temporal disorder on quantum phase transitions.
Naively, one might suspect that any time-dependent disorder (i.e., noise) destroys a quantum phase transition
because it acts as an effective temperature. However, it was recently shown that certain types of noise
preserve a quantum-critical state \cite{TDGA10} at least over a wide transient regime \cite{TDGA12}.
In cases in which a quantum phase transition survives, it is hard to see how it could escape the
stability criteria derived here. However, a detailed study of the applicability of our criteria to
quantum phase transitions remains a task for the future.

%%%%%%%%%%%%%%%%%%%%%%%%%%%%%%%%%%%%%%%%%%%%%%%%%%%%%%%%%%%%%%%%%%%%%%%%%%%%%%%%%
\section*{Acknowledgements}
%%%%%%%%%%%%%%%%%%%%%%%%%%%%%%%%%%%%%%%%%%%%%%%%%%%%%%%%%%%%%%%%%%%%%%%%%%%%%%%%%

This work was supported in part by the NSF under Grant Nos.\ DMR-1205803 and DMR-1506152.
T.V. acknowledges the hospitality of the Departamento de  F\'{i}sica,
Universidade Federal de Minas Gerais during the early stages of this work
as well as the SBF-APS Brazil-U.S. Professorship/Lectureship
Program. R.D. is grateful to CNPq, Brazil, for financial
support.

%%%%%%%%%%%%%%%%%%%%%%%%%%%%%%%%%%%%%%%%%%%%%%%%%%%%%%%%%%%%%%%%%%%%%%%%%%%%%%%%%
\appendix
%%%%%%%%%%%%%%%%%%%%%%%%%%%%%%%%%%%%%%%%%%%%%%%%%%%%%%%%%%%%%%%%%%%%%%%%%%%%%%%%%
\section{Correlation function of diffusive disorder}
\label{sec:Appendix_A}
%%%%%%%%%%%%%%%%%%%%%%%%%%%%%%%%%%%%%%%%%%%%%%%%%%%%%%%%%%%%%%%%%%%%%%%%%%%%%%%%%

The dynamics of the diffusive field $n(\mathbf{x},t)$ can be described by the Langevin
equation (see, e.g., Ref.\ \cite{HohenbergHalperin77})

\begin{equation}
\frac {\partial}{\partial t} n(\mathbf{x},t) = \lambda_0 \nabla^2 \frac {\partial H}{\partial n(\mathbf{x},t)} + \zeta(\mathbf{x},t)
\label{eq:Langevin_App}
\end{equation}
with Hamiltonian
\begin{equation}
H=\int d^dx\,  n^2(\mathbf{x},t)/(2C_0) - \mu \int d^dx\, n(\mathbf{x},t)~.
\label{eq:Hamiltonian}
\end{equation}
Here, $C_0$ equals the compressibility $\partial \langle n\rangle /\partial \mu$, and
$\zeta(\mathbf{x},t)$ is a conserving noise characterized by the correlation
function
\begin{equation}
\langle \zeta(\mathbf{x},t) \zeta(\mathbf{x'},t')  \rangle = -2 \Gamma\, \nabla^2 \delta(\mathbf{x}-\mathbf{x'}) \delta(t-t')
\label{eq:conserving noise}
\end{equation}
in real space or
\begin{equation}
\langle \zeta(\mathbf{q},t) \zeta(\mathbf{q'},t')  \rangle = 2 \Gamma\, \mathbf{q}^2 \delta(t-t') \delta(\mathbf{q}+\mathbf{q'})
\label{eq:conserving noise_q}
\end{equation}
in Fourier space.
Inserting the Hamiltonian (\ref{eq:Hamiltonian}) into the Langevin equation (\ref{eq:Langevin_App}),
we obtain the diffusion equation
\begin{equation}
\frac {\partial}{\partial t} n(\mathbf{x},t) - D \nabla^2 n(\mathbf{x},t) = \zeta(\mathbf{x},t)
\label{eq:diffusion}
\end{equation}
with diffusion constant $D=\lambda_0/C_0$. Using the Green function of the diffusion equation,
${\cal G}(\mathbf{q},t) = \exp(-D\mathbf{q}^2 t)$, we can write down a formal solution in Fourier space
(up to an additive constant),
\begin{equation}
n(\mathbf{q},t) = \int_{-\infty}^t dt' {\cal G}(\mathbf{q},t-t') \zeta(\mathbf{q},t')~.
\label{eq:formal_solution}
\end{equation}
The correlation function of $n(\mathbf{q},t)$ is now easily evaluated giving
\begin{equation}
G_{nn}(\mathbf{q},t-t')=\langle n(\mathbf{q},t) n(\mathbf{-q},t')\rangle = \frac \Gamma D e^{-D\mathbf{q}^2 |t-t'|}~.
\label{eq:n_corr_q}
\end{equation}
If the diffusive field is in thermal equilibrium at temperature $T$, it follows from (\ref{eq:Hamiltonian}) that
$\langle n(\mathbf{q},t) n(\mathbf{-q},t)\rangle= k_B T C_0 = k_B T (\partial \langle n \rangle/\partial \mu)$.
Therefore, $\Gamma/D= k_B T (\partial \langle n \rangle/\partial \mu)$.
Fourier transforming back to real space yields
\begin{equation}
G_{nn}(\mathbf{x},t) = \frac {k_B T (\partial \langle n \rangle/\partial \mu)}{(4\pi D|t|)^{d/2}} \exp\left[- \frac{\mathbf{x}^2}{4 D |t|} \right]~.
\label{eq:Gnn_App}
\end{equation}
This completes the derivation of eq.\ (\ref{eq:Gnn_diffusion}).

%%%%%%%%%%%%%%%%%%%%%%%%%%%%%%%%%%%%%%%%%%%%%%%%%%%%%%%%%%%%%%%%%%%%%%%%%%%%%%%%%
\section{Integrals leading to eqs.\ (\ref{eq:sigma_diffusive_late}) }
\label{sec:Appendix_B}
%%%%%%%%%%%%%%%%%%%%%%%%%%%%%%%%%%%%%%%%%%%%%%%%%%%%%%%%%%%%%%%%%%%%%%%%%%%%%%%%%

To calculate the integral (\ref{eq:sigmanbar_approx}) for the case of diffusive disorder,
we first perform the $x$-integration. As explained in the main text, this gives
$\int d^dx \, G_{nn}(\mathbf{x},t) = A$ for early times ($4Dt < (\xi/2)^2$) because
the integration range can be extended to infinity.
For late times, $4Dt > (\xi/2)^2$, we instead obtain $\int d^dx \, G_{nn}(\mathbf{x},t) =
A\xi^d/(4 \pi D |t|)^{d/2}$ because the exponential in $G_{nn}$ is approximately unity.

The remaining time integration in (\ref{eq:sigmanbar_approx}) covers both the early-time
and late-time regimes if $2 D \xi_t > (\xi/2)^2$. We therefore split the integration into two
parts, $\sigma^2_{\bar n} = \sigma^2_{\bar n, 1} + \sigma^2_{\bar n, 2}$ with
\begin{eqnarray}
\sigma^2_{\bar n, 1} &=& \frac {2} {\xi^d \xi_t} \int_0^{\xi^2/(16 D)} dt A = \frac {A} {8D} \xi_t^{-1} \xi^{2-d}~,
\label{eq:split_integral_1}\\
\sigma^2_{\bar n, 2} &=& \frac {2} {\xi^d \xi_t} \int_{\xi^2/(16 D)}^{\xi_t/2} dt \frac {A \xi^d}{(4\pi D t)^{d/2}}~.
\label{eq:split_integral_2}
\end{eqnarray}
The $\sigma^2_{\bar n, 2}$ integral depends on the dimensionality. For $d>2$, the integration range can be extended to infinity,
giving the leading behavior
\begin{equation}
\sigma^2_{\bar n, 2} \sim A D^{-1} \xi_t^{-1}\xi^{2-d}      \qquad  (d>2)~.
\label{eq:sig2a}
\end{equation}
The marginal case, $d=2$, gives a logarithm
\begin{equation}
\sigma^2_{\bar n, 2} \sim A D^{-1} \xi_t^{-1} \ln (8D\xi_t/\xi^2)      \qquad  (d=2)~.
\label{eq:sig2b}
\end{equation}
For $d<2$, the integral is dominated by its upper bound and yields
\begin{equation}
\sigma^2_{\bar n, 2} \sim A D^{-d/2} \xi_t^{-d/2}      \qquad  (d<2)~.
\label{eq:sig2c}
\end{equation}
Comparing the results (\ref{eq:sig2a}), (\ref{eq:sig2b}), (\ref{eq:sig2c}) to eq.\
(\ref{eq:split_integral_1}), we see that $\sigma^2_{\bar n, 2}$ is larger than $\sigma^2_{\bar n, 1}$ (in $d \le 2$)
or behaves the same as $\sigma^2_{\bar n, 1}$ (in $d>2$). $\sigma^2_{\bar n, 2}$ thus determines the final result
(\ref{eq:sigma_diffusive_late}).

\bibliographystyle{apsrev4-1}
\bibliography{../00Bibtex/rareregions}
\end{document}